\documentclass{aip-cp}

\usepackage[numbers]{natbib}
\usepackage{rotating}
\usepackage{subcaption}
\usepackage{graphicx}
\usepackage[utf8]{inputenc}

\newcommand\apj{ApJ}%
\newcommand\aap{A\&A}%
\newcommand\apjs{ApJS}%
\newcommand\na{New A}%
\def\mnras{MNRAS}%
\begin{document}

\title{Diffuse Gamma Rays in 3D Galactic Cosmic-ray Propagation Models}

\author[aff1]{R.~Kissmann\corref{cor1}}
\author[aff1]{F.~Niederwanger}
\author[aff1]{O.~Reimer}
\author[aff2]{A.~W.~Strong}

\affil[aff1]{Institut für Astro- und Teilchenphysik, Leopold Franzens Universität Innsbruck
Technikerstraße 25/8, \mbox{6020 Innsbruck}, Austria} 
\affil[aff2]{Max-Planck-Institut für extraterrestrische Physik, Garching, Germany}
\corresp[cor1]{Corresponding author: Ralf.Kissmann@uibk.ac.at}

\maketitle

\begin{abstract}
The \textsc{Picard} code for the numerical solution of the Galactic cosmic ray propagation problem allows for high-resolution models that acknowledge the 3D structure of our Galaxy. \textsc{Picard} was used to determine diffuse gamma-ray emission of the Galaxy over the energy range from 100\,MeV to 100\,TeV. 
We discuss the impact of a cosmic-ray source distribution aligned with the Galactic spiral arms for a range of such spiral-arm models.
As expected, the impact on the gamma-ray emission is most distinct in the  inverse-Compton channel, where imprints of the spiral arms are visible and yield predictions that are no longer symmetric to the rotational axis of the Milkyway. We will illustrate these differences by a direct comparison to results from previous axially symmetric Galactic propagation models: we find differences in the gamma-ray flux both on global scales and on local scales related to the spiral arm tangents.
We compare gamma-ray flux and spectra at on-arm vs. off-arm projections and characterize the differences to axially symmetric models.
\end{abstract}

\section{INTRODUCTION}
The fact that our Galaxy does not have an azimuthally symmetric shape is lately being acknowledged in recent Galactic cosmic-ray propagation models \citep[see, e.g.,][]{BenyaminEtAl2014ApJ782_34, EffenbergerEtAl2012AnA547A120, GaggeroEtAl2013PhRvL111_021102, JohannessonEtAl2013ICRC, WernerEtAl2015APh64_18, JohannessonEtAl2015ICRC}. These models focus on the observation that our Galaxy has a spiral structure \citep[see][]{Vallee2014ApJS215_1} and implications for the transport of cosmic rays within the Galaxy. This respective spiral structure will manifest itself in the distribution of ISM gas \citep[see][]{PohlEtAl2008ApJ677_283} and the interstellar magnetic field \citep[see, e.g.][]{FerriereTerral2014AnA561_100, Jansson2012ApJ757_14} but also in the distribution of cosmic-ray sources \citep[see, e.g.][]{BenyaminEtAl2014ApJ782_34, EffenbergerEtAl2012AnA547A120, Shaviv2003NewA8_39}.

The impact of the source distribution on the cosmic-ray flux and distribution within the Galaxy has been studied by several groups \citep[see][]{BenyaminEtAl2014ApJ782_34, EffenbergerEtAl2012AnA547A120, JohannessonEtAl2013ICRC, KoppEtAl2014NewA30_32, WernerEtAl2015APh64_18, KissmannEtAl2015APh70_39, JohannessonEtAl2015ICRC}. It was shown that a good agreement with the observed cosmic-ray data at Earth \citep[see][]{BenyaminEtAl2014ApJ782_34, KissmannEtAl2015APh70_39} can be found both in models with an axisymmetric source distribution and in models using a source distribution motivated by the observed three-dimensional structure of our Galaxy. Different spatial distributions of the sources lead to different spatial distributions of the cosmic-ray flux, which should lead to distinct effects on the emission of radiation by the interaction of cosmic rays with the interstellar medium resulting in the different models.

Here, we focus on the imprint of non-axisymmetric source distribution on the gamma-ray emission from our Galaxy. For this we use simulations performed with the \textsc{Picard} code. This code was introduced in \cite{Kissmann2014APh55_37} and is optimised for such spatially three-dimensional numerical modelling. In the following we introduce the \textsc{Picard} code with a focus on recent enhancements. Subsequently we discuss the ensuing gamma-ray emission for our models and show the related energy spectra.

\section{THE PICARD CODE}
The \textsc{Picard} code was introduced in \cite{Kissmann2014APh55_37} with a focus on spatially three-dimensional numerical modelling of Galactic cosmic-ray transport problems. The code is optimised to find steady-state solutions to the cosmic-ray transport equation:
\begin{equation}
\label{EqTransport}
  \frac{\partial \psi_i}{\partial t}
  =
  q(\vec{r}, p)
  +
  \nabla \cdot \mathcal{D}
  \nabla \psi_i
  +
  \frac{\partial}{\partial p} p^2 D_{pp}
  \frac{\partial}{\partial p} \frac{1}{p^2} \psi_i
  -
  \nabla\cdot \vec{v} \psi_i
  -
  \frac{\partial}{\partial p}
  \left\{
  \dot p \psi_i - \frac{p}{3} (\nabla \cdot\vec{v})\psi_i
  \right\}
  - 
  \frac{1}{\tau_f} \psi_i
  -
  \frac{1}{\tau_r} \psi_i,
\end{equation}
but it can also be applied to solve time-dependent problems. A numerical solution to the steady-state problem is found be discretising the differential operators in equation (\ref{EqTransport}) in space and momentum and then finding an approximate solution to the resulting linear system of equations. In \textsc{Picard} the linear system of equations can be solved using different standard techniques, including a range of multigrid methods \citep[see][]{TrottenbergEtAlBook2001} or the BiCGStab method as described in \cite{sleijpen1993bicgstab}, with the possibility to add other solvers into the modular framework of the code.

\textsc{Picard} features all relevant transport processes inherent in equation (\ref{EqTransport}): spatial diffusion, spatial convection, momentum diffusion and relevant energy-loss processes. Fluxes of light species depend on those of heavier species due to the possible fragmentation or radiative-decay processes. This is realised through the fragmentation and radiative-decay timescales $\tau_f$ and $\tau_r$, respectively, together with a consideration of a related source term for other species. The resulting nuclear network is efficiently implemented in \textsc{Picard}, leading to a numerically optimised and accurate treatment of coupled nuclear species \citep[see][]{KissmannEtAl2015APh70_39}.

Prominent features of the \textsc{Picard} code include the possibility to consider anisotropic spatial diffusion \citep[for a discussion see, e.g.][]{EvoliEtAl2012PhRvL108_1102, Effenberger2012ApJ750_108} and arbitrary convection velocities $\vec{v}$. The numerical solver is MPI-parallel, thus facilitating the possibility to achieve up to deca-parsec scale resolution by making use of large-scale distributed-memory computers. The numerical solution computed using \textsc{Picard} is acquired with a user-defined accuracy without the need to check the quality of the numerical solution a-posteriory. \textsc{Picard} is continuously expanded where, e.g.,  the currently ongoing implementation of an improved interstellar radiation field is discussed by \cite{NiederwangerEtAl2016AIPC}.

With this code we studied the implications of spiral-arm cosmic-ray source distributions on the spatial variability of the cosmic-ray flux in \cite{WernerEtAl2015APh64_18}. Additionally, we showed in \cite{KissmannEtAl2015APh70_39} that the transport parameters in equation (\ref{EqTransport}) can be adapted to such localised source distributions, thus leading to a good agreement between predicted cosmic-ray fluxes at Earth and corresponding observations. For the secondary to primary ratios there appears a dependence on the distance to the nearest sources in modules using such localised sources. In the following we will address the impact of these models on the Galactic diffuse gamma-ray emission.

\section{GAMMA-RAY EMISSION}
\label{SecMaps}
In \cite{KissmannEtAl2015APh70_39} we found that a Galactic
cosmic-ray propagation model featuring localised sources can reproduce
cosmic-ray measurements at Earth as well as a model with underlying
axisymmetry. In contrast to such an axisymmetric model, a spiral-arm
source distribution leads to a localisation of the cosmic-ray
flux, at least for the primary cosmic-ray species. The distinct
difference in spatial distribution between the different models can be
expected to lead to differences in the resulting gamma-ray flux in the
different models. In the following we focus on the comparison between a model using an axisymmetric source distribution and one using a spiral-arm source distribution with four spiral arms. Details can be found in \cite{WernerEtAl2015APh64_18} and \cite{KissmannEtAl2015APh70_39}, where this specific model is referred to as the \emph{Steiman-model}.

\begin{figure}
\begin{tabular}{c}
\includegraphics[width=0.5\textwidth]{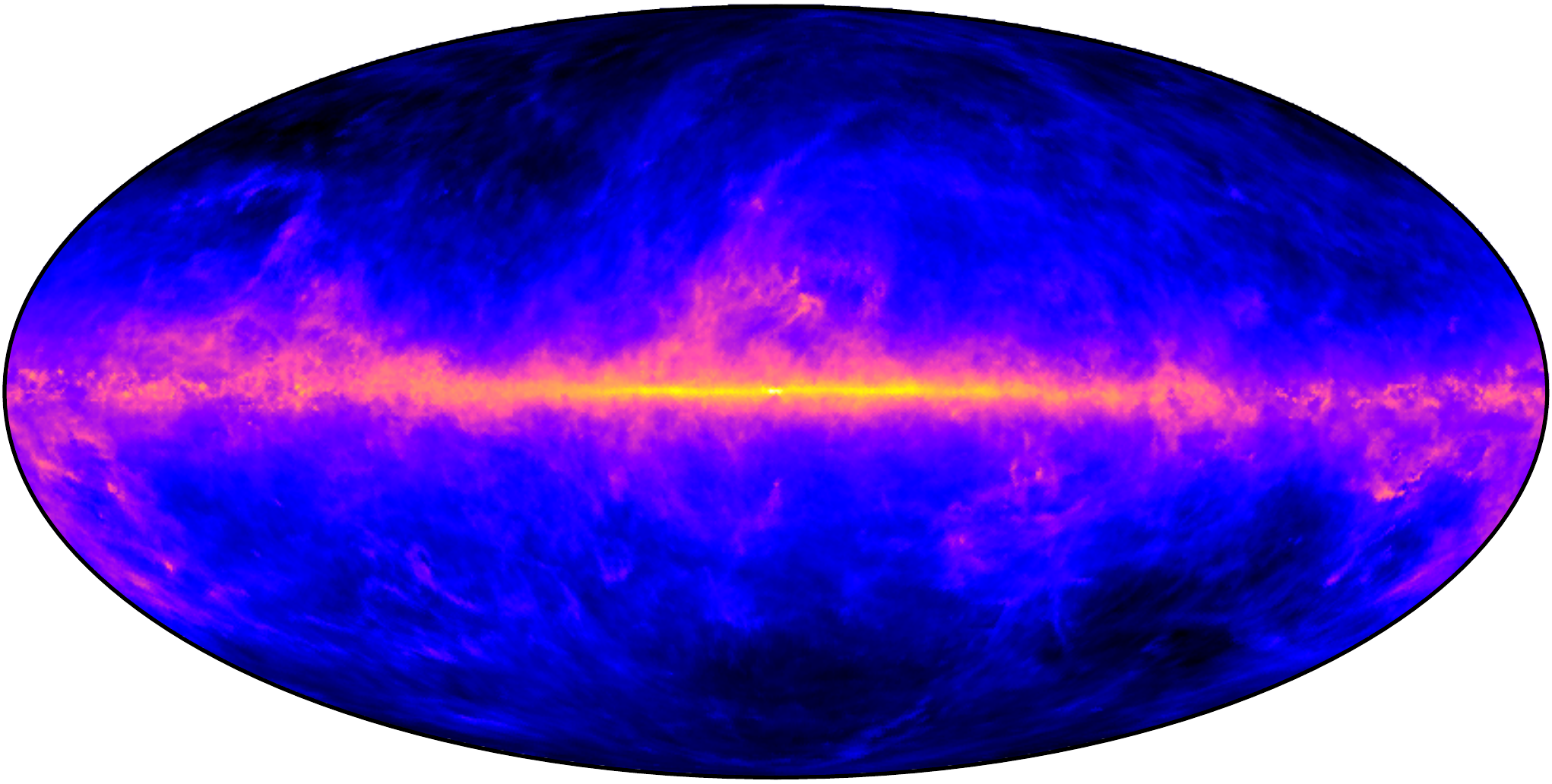}
\\
\includegraphics[width=0.5\textwidth]{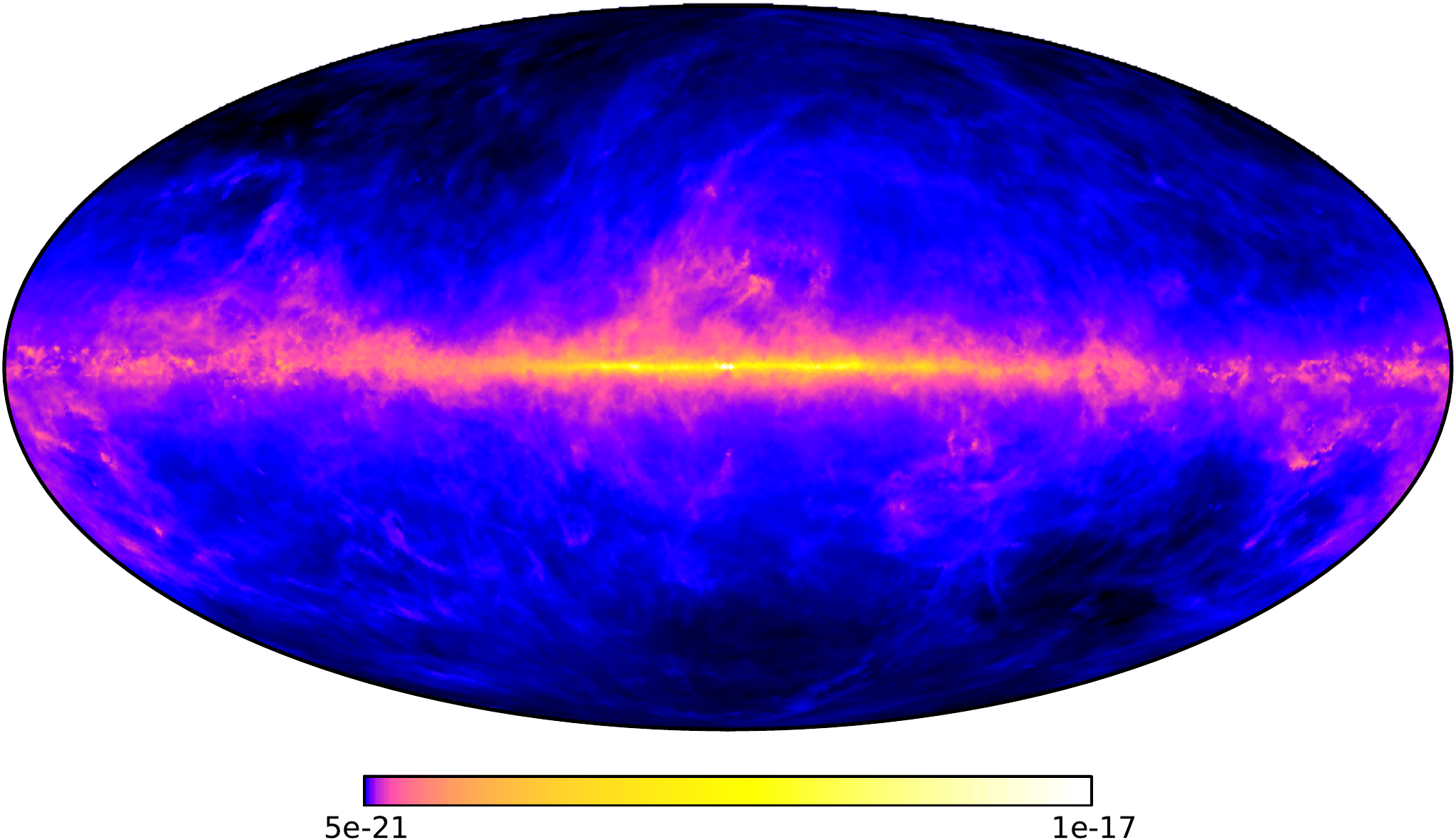}
\end{tabular}
\caption{\label{FigGammaEmissTotal}All-sky gamma-ray emission at $E_{\gamma} = 10$~TeV for an axisymmetric (top) and a four-arm (bottom) source distribution. The gamma-ray emission is shown in units of cm$^{-2}$ s$^{-1}$ sr$^{-1}$ MeV$^{-1}$.}
\end{figure}

The relevant channels for diffuse gamma-ray emission are $\pi^0$-decay, bremsstrahlung, and inverse Compton (IC) radiation. The emissivity of the former two of these processes results from the product of the gas-density with the respective cosmic-ray density -- in this case nucleons and electrons, respectively. Therefore, it is not easy to distinguish whether a spatial feature in the gamma-ray emission results from local structures in the gas distribution or the cosmic ray distribution. Since current models of the Galactic radiation field are spatially much smoother than the gas distribution \citep[see][]{StrongEtAl2000ApJ537_763S, PorterEtAl2008ApJ682_400, PopescuTuffs2013MNRAS436_1302}, the IC channel is a much clearer tracer of the localisation of the sources within our models. This is obvious when investigating the gamma-ray emission in Fig. \ref{FigGammaEmissTotal}. A comparison shows that the Galactic plane near the Galactic centre becomes slightly brighter for the model with a spiral-arm source distribution. Otherwise localised features affected by the different source models are not readily apparent.

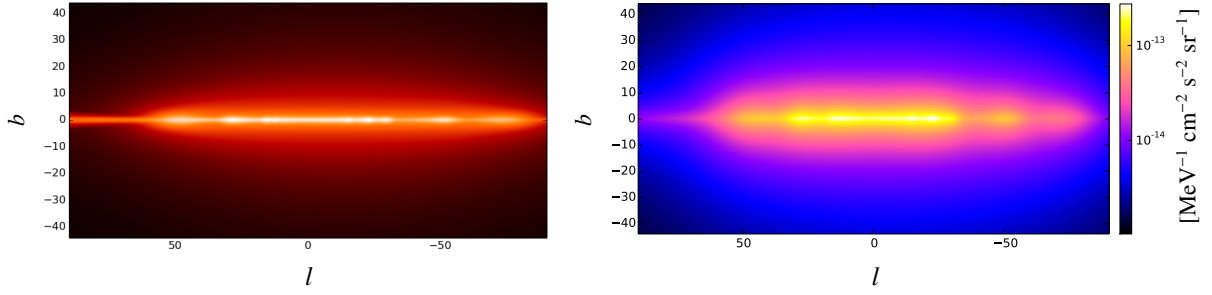
\begin{figure}
\centerline{
\setlength{\unitlength}{0.0002\textwidth}
	\begin{picture}(2248,1200)(-200,-200)
	\put(1060,-150){$l$}
	\put(-150,500){\small \rotatebox{90}{$b$}}
	\includegraphics[height=1000\unitlength]{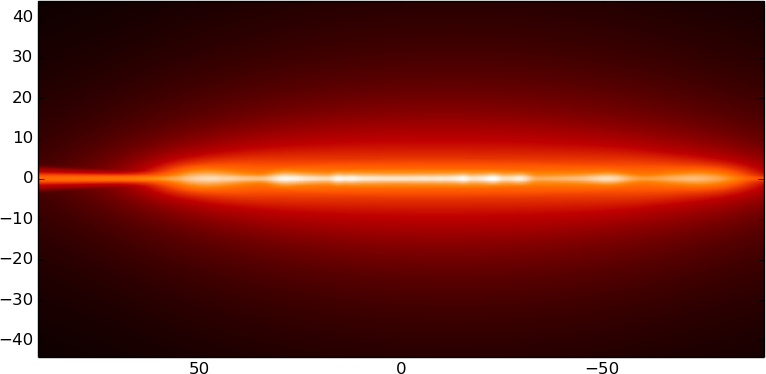}
	\end{picture}
	~\hfill
	\begin{picture}(2651,1200)(-200,-200)
	\put(1060,-150){$l$}
	\put(-150,500){\small \rotatebox{90}{$b$}}
	\put(2300,100){\small \rotatebox{90}{[MeV$^{-1}$ cm$^{-2}$ s$^{-2}$ sr$^{-1}$]}}
	\includegraphics[height=1000\unitlength]{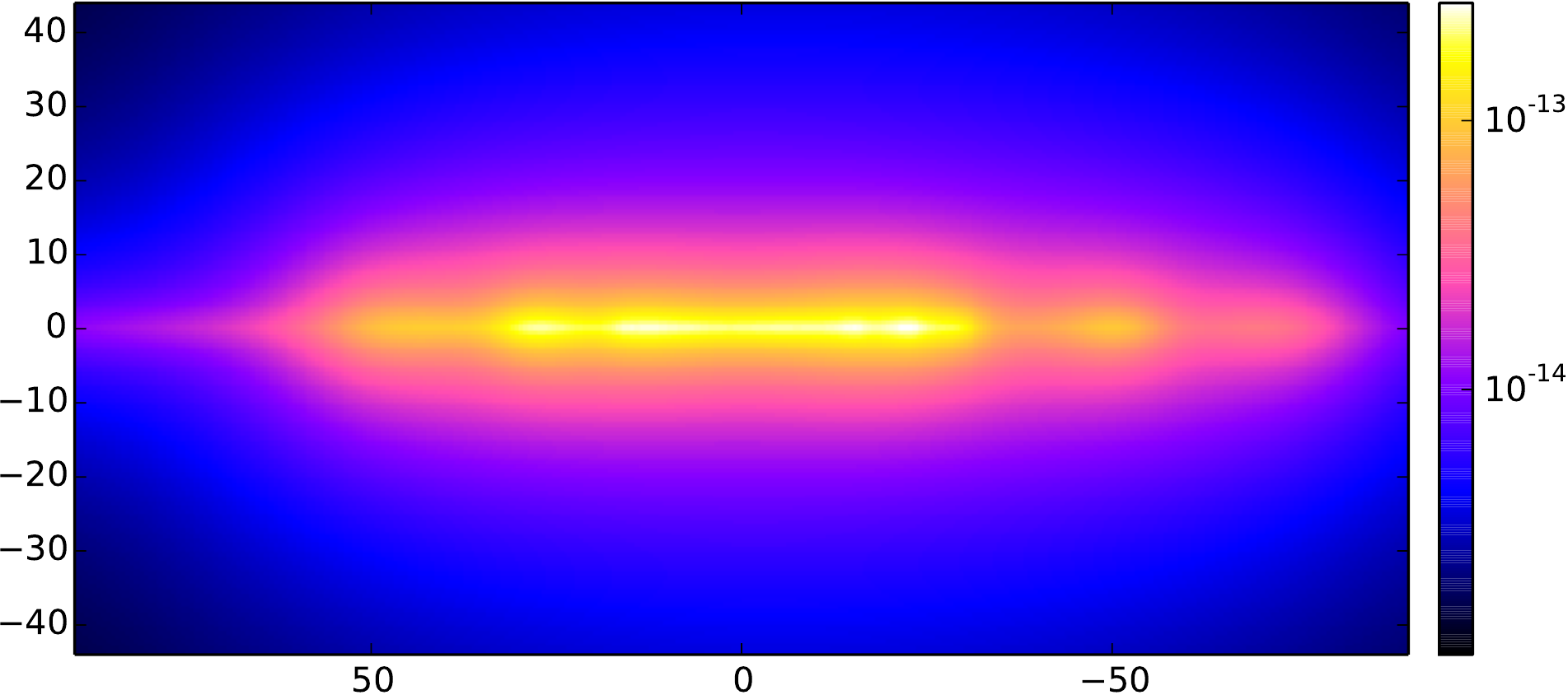}
	\end{picture}
	}
	\caption{\label{FigSourceProjection}Left: Projected source intensity for the four-arm source-distribution model. Spiral-arm tangents are visible as enhanced intensity in the Galactic plane. Right: Inverse Compton gamma-ray emission at 100~GeV for the same model.}
\end{figure}

In the following we will focus on the region near the Galactic centre, where the localised features of the cosmic-ray distribution in the Steiman-model have the most obvious impact on the gamma-ray emission. This is caused by the spiral arms that are tightly wound around the Galactic centre leading to several spiral arm tangents near the Galactic centre along which the projected source intensity is enhanced. This is shown in Fig. \ref{FigSourceProjection} where the projected source intensity near the Galactic centre region is shown. The specific spiral-arm tangents are consistent with observations \citep[][]{Vallee2014ApJS215_1} since the Steiman model is driven by observations of the Galactic spiral arms \citep[][]{Steiman-CameronEtAl2010ApJ722_1460}. The specific spiral-arm tangents visible in Fig. \ref{FigSourceProjection} are tangents to the Sagittarius / Carina arm at $l\sim$50$^{\circ}$, $l\sim$-16$^{\circ}$, and $l\sim$-75$^{\circ}$, to the Scutum / Crux Centaurus arm at $l\sim$30$^{\circ}$ and $l\sim$-50$^{\circ}$, the Norma arm at $l\sim$19$^{\circ}$ and at $l\sim$-30$^{\circ}$, and the Perseus arm at $l\sim$-22$^{\circ}$.

In Fig. \ref{FigSourceProjection} we also show the IC emission at an energy of $\sim$100~GeV. The increase in gamma-ray flux at the positions of the spiral-arm tangents is obvious at this energy. High-energy electrons are tightly coupled to their sources due to their very high energy losses \citep[][]{WernerEtAl2015APh64_18}. Therefore, the related gamma-ray emission traces these source regions leading to an imprint of the spiral-arm tangents also on the gamma-ray emission. In the total gamma-ray emission these effects are not so obvious, because of the impact of the gas-distribution discussed above and also because the Galactic diffuse gamma-ray emission is in many directions dominated by $\pi^0$-decay emission. 
In the following we investigate the quantitative impact of the spiral-arm source distribution on the gamma-ray emission.

\section{GAMMA-RAY SPECTRA}
\label{SecSpectra}
\begin{figure}
\centerline{
\setlength{\unitlength}{0.0003\textwidth}
	\begin{picture}(1554,1150)(-150,-150)
	\put(650,-100){$E$ [MeV]}
	\put(-100,240){\small \rotatebox{90}{$J_{\gamma} $[MeV cm$^{-2}$ s$^{-2}$ sr$^{-1}$]}}
	\includegraphics[height=1000\unitlength]{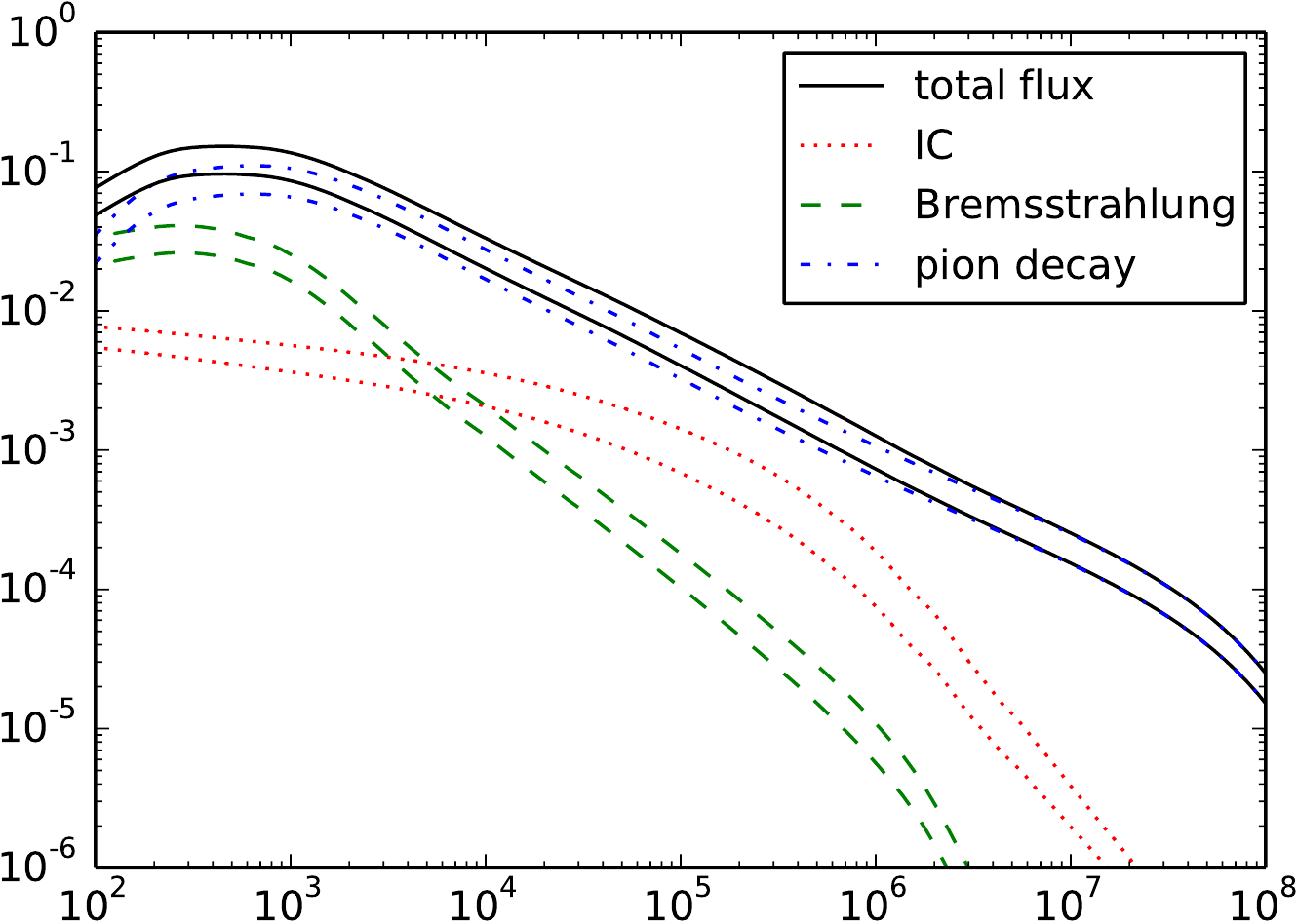}
	\end{picture}
	\hfill
	\begin{picture}(1554,1150)(-150,-150)
	\put(650,-100){$E$ [MeV]}
	\put(-100,240){\small \rotatebox{90}{$J_{\gamma} $[MeV cm$^{-2}$ s$^{-2}$ sr$^{-1}$]}}
	\includegraphics[height=1000\unitlength]{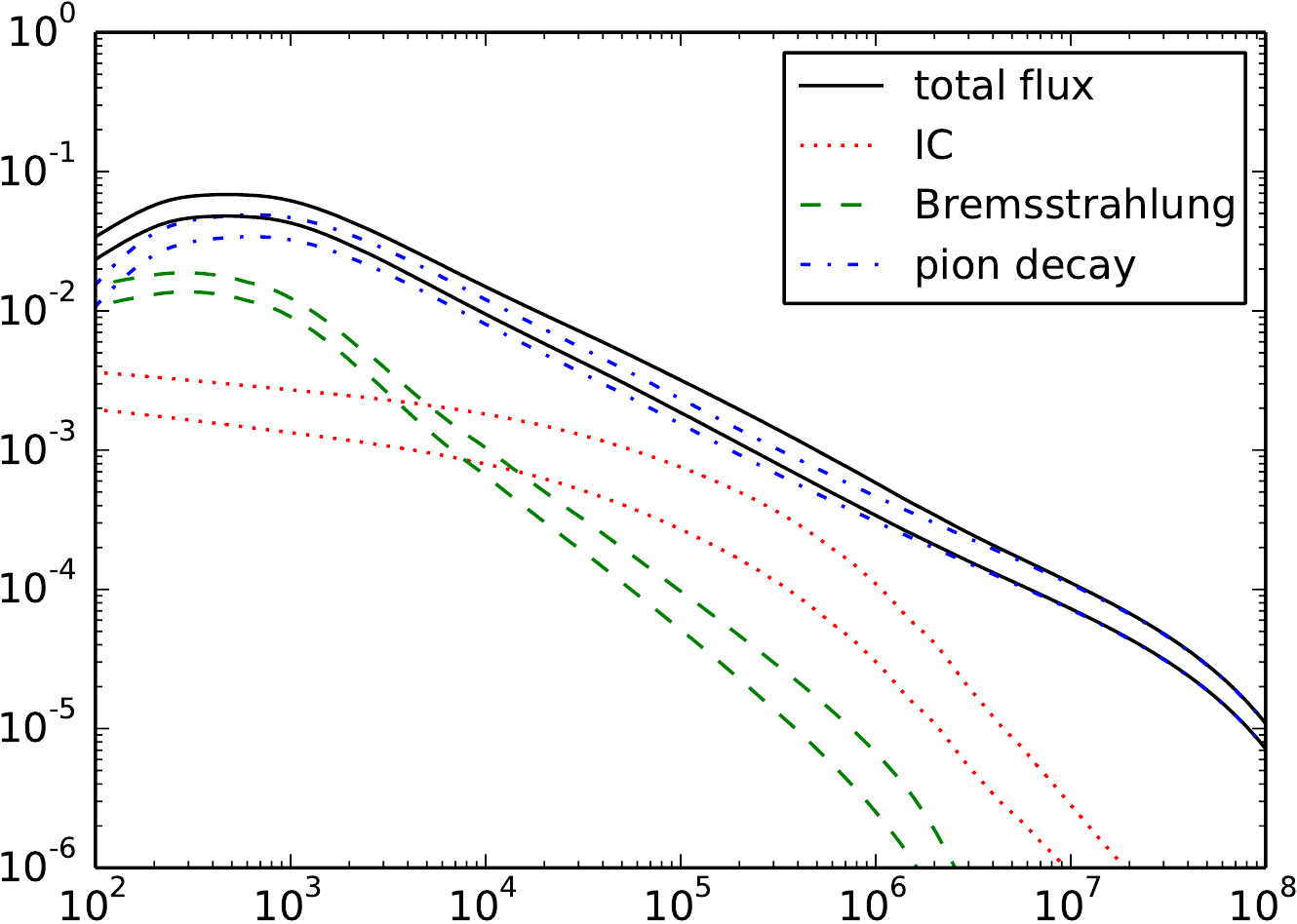}
	\end{picture}
}
\caption{\label{FigSpectra}Spectra for the the different gamma-ray emission channels for two of the spiral-arm tangents. In each plot on-arm and off-arm data are compared, where the on-arm spectra are the ones with the higher fluxes. On the left results are shown for a tangent of the Norma arm with the on-arm data taken at -30$^{\circ}$ longitude and the off-arm data taken at -35$^{\circ}$ longitude. On the right corresponding data for a Sagittarius-arm tangent are shown with the on-arm data taken at 50$^{\circ}$ longitude and the off-arm data taken at 62$^{\circ}$ longitude. Results are integrated over a circle with a radius of 2$^{\circ}$.}
\end{figure}

For a quantitative analysis we compare gamma-ray spectra at on-arm and off-arm positions. In Fig. \ref{FigSpectra} corresponding results are shown for spiral-arm tangents of the Norma and the Sagittarius arm. Obviously, the on-arm flux is significantly higher in both cases and for all gamma-ray emission channels. Especially, for Bremsstrahlung and IC the on-arm spectra are also harder. For these channels we found that the power-law index can increase by 0.1 or even more at a spiral-arm tangent as compared to an off-arm position. This is somewhat more pronounced for the Sagittarius arm, because in the case of the Norma arm the off-arm region is in a region featuring several other spiral-arms. Thus, the Norma off-arm region is contaminated by the presence of other close-by cosmic-ray sources.


The spectral hardening of the IC flux can be explained by the energy dependence of the electron's energy losses. Due to their rapid increase with energy, high-energy electrons suffer from the most severe energy losses. Therefore, high-energy electrons can only be found in the close vicinity of their sources. In \cite{WernerEtAl2015APh64_18} a significant hardening of the electron spectrum was observed for on-arm positions. This naturally translates into a hardening of the resulting photon spectrum for a sight line in the direction of a spiral-arm tangent.

The $\pi^0$ channel shows only mild change in spectral slope. This is also consistent with the observations in \cite{WernerEtAl2015APh64_18} where the proton flux was found to mainly change in amplitude from an on-arm to an off-arm position. This reflects the fact that the spatial distribution of nuclear cosmic rays is mainly driven by spatial diffusion due to their rather low energy losses. Also in gamma rays this channel mainly features a change in amplitude when comparing an on-arm and an off-arm region.

Since the total flux in both cases is dominated by $\pi^0$ decay, the change in slope for the total flux is also rather small. Only in the energy regime from a few up to a few hundred GeV, where the contribution of IC emission to the total gamma-ray flux is largest, the spectrum at an on-arm position becomes harder with a change in power law index of up to 0.05. In the present case we used the same source-distribution for electrons and nuclei. In a model using different source distributions the corresponding changes would become larger, for such localised source distributions as were discussed in this case. In fact, for high-energy electrons it will be important to consider the actual sources \citep[see][]{Fermi2015ApJS218_23,Fermi2016ApJS222_5,AharonianEtAl2006ApJ636_777} when aiming for an accurate model for the diffuse gamma-ray emission in the Galactic plane. Our model shows that localised sources will not only have an impact on the gamma-ray emission in the Galactic plane, but can also affect the local spectral slope. This will make a template-fitting approach at least more problematic than a physically motivated modelling approach based on the propagation physics of Galactic cosmic rays. One should be aware that in the current approach only the distribution of the cosmic-ray sources was adapted to the Galactic structure. With an adoption of the gas-, magnetic field-, and possibility the interstellar radiation distribution the differences between an axisymmetric approach and one acknowledging the actual structure can become even larger, holding the possibility to also explain the small-scale structure in the gamma-ray emission in the Galactic plane.

\section{CONCLUSIONS}
In this study we showed that \textsc{Picard} is ready to predict gamma-ray fluxes for a comparison with data in the context of Galactic cosmic-ray transport models taking into account the three-dimensional structure of our Galaxy.
 We showed a first quantitative discussion of the impact of localised sources on the Galactic diffuse gamma-ray emission. We found a hardening of the gamma-ray emission spectrum in the direction of the cosmic-ray sources for the leptonic emission channels. Since in our model the emission is dominated by $\pi^0$ decay the change in slope for the total gamma-ray flux is rather small and also limited to energies between 1 and a few hundred GeV, where the leptonic channels provide a comparatively large fraction of the total emission. The current analysis was focussed on the impact of the source distribution. While this can only be viewed as a first step, with the future need to include the full Galactic structure for gas-, magnetic field-, and interstellar radiation-distribution, only a dissemination of the individual effects can show the respective impact on changes in the gamma-ray emission.



\section{ACKNOWLEDGMENTS}
The computational results presented have been achieved (in part) using
the HPC infrastructure LEO and MACH of the University of Innsbruck. \\
FN acknowledges financial support from the Austrian Science Fund (FWF) project number
\textit{I1345}, in collaboration with the French Science Fund (ANR),
project ID \textit{ANR-13-IS05-0001}.





\begin{thebibliography}{10}

\bibitem{BenyaminEtAl2014ApJ782_34}
D.~{Benyamin}, E.~{Nakar}, T.~{Piran}, et~al.
\newblock {Recovering the Observed B/C Ratio in a Dynamic Spiral-armed Cosmic
  Ray Model}.
\newblock {\em \apj\/}, 782:34, February 2014.

\bibitem{EffenbergerEtAl2012AnA547A120}
F.~{Effenberger}, H.~{Fichtner}, K.~{Scherer}, et~al.
\newblock {Anisotropic diffusion of Galactic cosmic ray protons and their
  steady-state azimuthal distribution}.
\newblock {\em \aap\/}, 547:A120, November 2012.

\bibitem{GaggeroEtAl2013PhRvL111_021102}
D.~{Gaggero}, L.~{Maccione}, G.~{Di Bernardo}, et~al.
\newblock {Three-Dimensional Model of Cosmic-Ray Lepton Propagation Reproduces
  Data from the Alpha Magnetic Spectrometer on the International Space
  Station}.
\newblock {\em Physical Review Letters\/}, 111(2):021102, July 2013.

\bibitem{JohannessonEtAl2013ICRC}
G.~{J\'ohannesson}, I.~V. {Moskalenko}, and T.~{Porter}.
\newblock {Toward 3D mapping of the interstellar medium in the Milky Way:
  impact on cosmic rays and diffuse emission }.
\newblock {\em Proceedings of the 33rd International Cosmic Ray Conference\/},
  2013.

\bibitem{WernerEtAl2015APh64_18}
M.~{Werner}, R.~{Kissmann}, A.~W. {Strong}, et~al.
\newblock {Spiral arms as cosmic ray source distributions}.
\newblock {\em Astroparticle Physics\/}, 64:18--33, April 2015.

\bibitem{JohannessonEtAl2015ICRC}
G.~{J\'ohannesson}, I.~V. {Moskalenko}, E.~{Orlando}, et~al.
\newblock {The Effects of Three Dimensional Structures on Cosmic-Ray
  Propagation and Interstellar Emissions}.

\bibitem{Vallee2014ApJS215_1}
J.~P. {Vall{\'e}e}.
\newblock {Catalog of Observed Tangents to the Spiral Arms in the Milky Way
  Galaxy}.
\newblock {\em \apjs\/}, 215:1, November 2014.

\bibitem{PohlEtAl2008ApJ677_283}
M.~{Pohl}, P.~{Englmaier}, and N.~{Bissantz}.
\newblock {Three-Dimensional Distribution of Molecular Gas in the Barred Milky
  Way}.
\newblock {\em \apj\/}, 677:283--291, April 2008.

\bibitem{FerriereTerral2014AnA561_100}
K.~{Ferri{\`e}re} and P.~{Terral}.
\newblock {Analytical models of X-shape magnetic fields in galactic halos}.
\newblock {\em \aap\/}, 561:A100, January 2014.

\bibitem{Jansson2012ApJ757_14}
R.~{Jansson} and G.~R. {Farrar}.
\newblock {A New Model of the Galactic Magnetic Field}.
\newblock {\em \apj\/}, 757:14, September 2012.

\bibitem{Shaviv2003NewA8_39}
N.~J. {Shaviv}.
\newblock {The spiral structure of the Milky Way, cosmic rays, and ice age
  epochs on Earth}.
\newblock {\em \na\/}, 8:39--77, January 2003.

\bibitem{KoppEtAl2014NewA30_32}
A.~{Kopp}, I.~{B{\"u}sching}, M.~S. {Potgieter}, et~al.
\newblock {A stochastic approach to Galactic proton propagation: Influence of
  the spiral arm structure}.
\newblock {\em \na\/}, 30:32--37, July 2014.

\bibitem{KissmannEtAl2015APh70_39}
R.~{Kissmann}, M.~{Werner}, O.~{Reimer}, et~al.
\newblock {Propagation in 3D spiral-arm cosmic-ray source distribution models
  and secondary particle production using PICARD}.
\newblock {\em Astroparticle Physics\/}, 70:39--53, October 2015.

\bibitem{Kissmann2014APh55_37}
R.~{Kissmann}.
\newblock {PICARD: A novel code for the Galactic Cosmic Ray propagation
  problem}.
\newblock {\em Astroparticle Physics\/}, 55:37--50, March 2014.

\bibitem{TrottenbergEtAlBook2001}
U.~{Trottenberg}, , C.~W. {Ooosterlee}, et~al.
\newblock {\em Multigrid\/}.
\newblock Academic Press, Inc., Orlando, FL, USA, 2001.

\bibitem{sleijpen1993bicgstab}
Gerard~LG Sleijpen and Diederik~R Fokkema.
\newblock Bicgstab (l) for linear equations involving unsymmetric matrices with
  complex spectrum.
\newblock {\em Electronic Transactions on Numerical Analysis\/}, 1(11):2000,
  1993.

\bibitem{EvoliEtAl2012PhRvL108_1102}
C.~{Evoli}, D.~{Gaggero}, D.~{Grasso}, et~al.
\newblock {Common Solution to the Cosmic Ray Anisotropy and Gradient Problems}.
\newblock {\em Physical Review Letters\/}, 108(21):211102, May 2012.

\bibitem{Effenberger2012ApJ750_108}
F.~{Effenberger}, H.~{Fichtner}, K.~{Scherer}, et~al.
\newblock {A Generalized Diffusion Tensor for Fully Anisotropic Diffusion of
  Energetic Particles in the Heliospheric Magnetic Field}.
\newblock {\em \apj\/}, 750:108, May 2012.

\bibitem{NiederwangerEtAl2016AIPC}
F.~{Niederwanger}, R.~{Kissmann}, O.~{Reimer}, et~al.
\newblock {The use case of a new Interstellar Radiation Field for Diffuse
  Galactic Gamma-ray Emission Models}.
\newblock In F.~A. {Aharonian}, W.~{Hofmann}, and F.~M. {Rieger}, editors, {\em
  American Institute of Physics Conference Series\/}, American Institute of
  Physics Conference Series. 2016.

\bibitem{StrongEtAl2000ApJ537_763S}
A.~W. {Strong}, I.~V. {Moskalenko}, and O.~{Reimer}.
\newblock {Diffuse Continuum Gamma Rays from the Galaxy}.
\newblock {\em \apj\/}, 537:763--784, July 2000.

\bibitem{PorterEtAl2008ApJ682_400}
T.~A. {Porter}, I.~V. {Moskalenko}, A.~W. {Strong}, et~al.
\newblock {Inverse Compton Origin of the Hard X-Ray and Soft Gamma-Ray Emission
  from the Galactic Ridge}.
\newblock {\em \apj\/}, 682:400-407, July 2008.

\bibitem{PopescuTuffs2013MNRAS436_1302}
C.~C. {Popescu} and R.~J. {Tuffs}.
\newblock {Radiation fields in star-forming galaxies: the disc, thin disc and
  bulge}.
\newblock {\em \mnras\/}, 436:1302--1321, December 2013.

\bibitem{Steiman-CameronEtAl2010ApJ722_1460}
T.~Y. {Steiman-Cameron}, M.~{Wolfire}, and D.~{Hollenbach}.
\newblock {COBE and the Galactic Interstellar Medium: Geometry of the Spiral
  Arms from FIR Cooling Lines}.
\newblock {\em \apj\/}, 722:1460--1473, October 2010.

\bibitem{Fermi2015ApJS218_23}
F.~{Acero}, M.~{Ackermann}, M.~{Ajello}, et~al.
\newblock {Fermi Large Area Telescope Third Source Catalog}.
\newblock {\em \apjs\/}, 218:23, June 2015.

\bibitem{Fermi2016ApJS222_5}
M.~{Ackermann}, M.~{Ajello}, W.~B. {Atwood}, et~al.
\newblock {2FHL: The Second Catalog of Hard Fermi-LAT Sources}.
\newblock {\em \apjs\/}, 222:5, January 2016.

\bibitem{AharonianEtAl2006ApJ636_777}
F.~{Aharonian}, A.~G. {Akhperjanian}, A.~R. {Bazer-Bachi}, et~al.
\newblock {The H.E.S.S. Survey of the Inner Galaxy in Very High Energy Gamma
  Rays}.
\newblock {\em \apj\/}, 636:777--797, January 2006.

\end{thebibliography}
\end{document}